\documentclass[a4paper]{article}

\usepackage{amsmath}
\usepackage{graphicx}
\begin{document}

\begin{titlepage}
\begin{center}

\LARGE{\textbf{Analytic Amplitudes for Hadronic Forward Scattering and the Heisenberg $\ln^2s$ Behaviour of Total Cross Sections}\footnote{Invited talk at the QCD International Conference QCD03, Montpellier, France, July 2-9, 2003.}}

\vspace{1cm}

\Large{\textbf{Basarab Nicolescu}}

\vspace{1cm}

\large{\textsl{LPNHE (Unit\'e de Recherche des Universit\'es
Paris 6 et Paris 7, associ\'ee au CNRS), Theory Group,
Universit\'e Pierre et Marie Curie, 4 Place Jussieu, 75252 Paris Cedex 05, France}}

\vspace{0.2cm}

\textsl{email: nicolesc@lpnhep.in2p3.fr}

\end{center}

\vspace{3cm}

\begin{center}
\textbf{Abstract}

The $\ln^2s$ behaviour of total cross sections, first obtained by Heisenberg 50 years ago, receives now increased interest both on phenomenological and theoretical levels.We present a modification of the Heisenberg's model in connection with the presence of glueballs and we show that it leads to a realistic description of all existing hadron total cross-sections data, in agreement with the COMPETE analysis.
\end{center}

\end{titlepage}

\section{The COMPETE analysis of forward data}
Analytic parametrizations of forward $(t=0)$ hadron scattering 
amplitudes is a well-established domain in strong interactions. 

However, in the past, the phenomenology of forward scattering had 
quite a high degree of arbitrariness :
i) An excessive focus on $pp$ and $\bar pp$ scattering ; 
ii) Important physical constraints are often mixed with less general 
    or even ad-hoc properties ;
iii) The cut-off in energy, defining the region of applicability of the 
    high-energy models, differs from one author to the other ;
iv) The set of data considered by different authors is sometimes not 
    the same ;
v) No rigorous connection is made between the number of parameters 
    and the number of data points ;
 vi) No attention was paid to the necessity of the stability of 
    parameter values ;
vii) The experiments were performed in the past in quite a chaotic 
    way : huge gaps are sometimes present between low-energy and 
    high-energy domains or inside the high-energy domain itself.

The COMPETE (\underline{CO}mputerized \underline{M}odels and 
\underline{P}arameter \underline{E}valuation for \underline{T}heo\-ry 
and \underline{E}xpe\-riment) project tries to cure 
as much as possible the above discussed arbitrariness.

The $\chi^2/dof$ criterium is not able, 
by itself, to cure the above difficulties : new indicators have 
to be defined.
Once these indicators are defined\cite{Cudell}, an appropriate sum of their 
numerical values is proposed in order to establish the \textit{rank}
of the model under study : the highest the numerical value of the rank 
the better the model under consideration.

The final aim of the COMPETE project is to provide our community with a 
periodic cross assessments of data and models via computer-readable 
files on the Web \cite{web}.

We consider the following exemplar cases of the imaginary part of the 
scattering amplitudes :
\begin{equation}
    ImF^{ab}=s\sigma_{ab}=P_{1}^{ab}+P_{2}^{ab}+R_{+}^{ab}
    \pm R_{-}^{ab}
    \label{eq:1prime}
\end{equation}
where : \\
- the $\pm$ sign in formula (\ref{eq:1prime}) corresponds to 
antiparticle (resp. particle) - particle scattering amplitude.\\
- $R_{\pm}$ signify the effective secondary-Reggeon 
($(f,a_{2}),\ (\rho,\omega)$) contributions to the 
even (odd)-under-crossing amplitude
\begin{equation}
    R_{\pm}(s)=Y_{\pm}\left(\frac{s}{s_{1}}\right)^{\alpha_{\pm}},
    \label{eq:2}
\end{equation}
where $Y$ is a constant residue, $\alpha$ - the reggeon intercept and 
$s_{1}$ - a scale factor fixed at 1 GeV$^2$ ;\\
- $P_{1}(s)$ is the contribution of the Pomeron Regge pole
\begin{equation}    
    P_{1}^{ab}(s)=C_{1}^{ab}\left(\frac{s}{s_{1}}\right)^{\alpha_{P_{1}}},
    \label{eq:3}
\end{equation}
$\alpha_{P_{1}}$ is the Pomeron intercept
$
    \alpha_{P_{1}}=1,
$
and $C^{ab}$ are constant residues.\\
- $P_{2}^{ab}(s)$ is the second component of the Pomeron 
corresponding to three different $J$-plane singularities :
\begin{itemize}
\item[a)] a Regge simple - pole contribution
    \begin{equation}    
    P_{2}^{ab}(s)=C_{2}^{ab}\left(\frac{s}{s_{1}}\right)^{\alpha_{P_{2}}}, 
    \label{eq:5}
    \end{equation}    
with
    $\alpha_{P_{2}}=1+\epsilon,\  \epsilon >0,$
    and $C_{2}^{ab}$ const. ;
     
    \item  [b)] a Regge double-pole contribution
    \begin{equation}                        
        P_{2}^{ab}(s)=s\left[A^{ab}+B^{ab}\ln\left(\frac{s}{s_{1}}
        \right)\right], 
    \label{eq:7}
    \end{equation}      
   with $A^{ab}$ and $B^{ab}$ const. ;
      
    \item  [c)] a Regge triple-pole contribution
    \begin{equation}
        P_{2}^{ab}(s)=s\left[A^{ab}+B^{ab}\ln^2\left(\frac{s}{s_{0}}
        \right)\right],
        \label{eq:8}
    \end{equation}
where $A^{ab}$ and $B^{ab}$ are constants and $s_{0}$ is an arbitrary 
scale factor.
\end{itemize}
We consider all the existing forward data for $pp,\bar pp, \pi p, Kp, 
\gamma\gamma$ and $\Sigma p$ scatterings.
The number of data points is : 904, 742, 648,  569, 498, 453, 397, 
329 when the cut-off in energy is 3, 4, 5, 6, 7, 8, 9, 10 GeV 
respectively.
A large number of variants were studied and classified. 
All definitions and numerical details can be found in Ref. 1.

The 2-component Pomeron classes of models are
    RRPE, RRPL and RRPL2,
where by RR we denote the two effective secondary-reggeon 
contributions, by P - the contribution of the Pomeron Regge-pole 
located at $J=1$, by E - the contribution of the Pomeron Regge-pole 
located at $J=1+\epsilon$, by L - the contribution of the component of 
the Pomeron, located at $J=1$ (double pole), and by L2 - the contribution of the 
component of the Pomeron located at $J=1$ (triple pole).
We also studied the 1-component Pomeron classes of models
    RRE, RRL and RRL2. 
  
The highest rank are get by the RRPL2$_u$ mo\-dels (see Table 1 and Figure 1), corresponding to the $\ln^2s$ behaviour of total cross sections first proposed by Heisenberg 50 years ago \cite{Heisen52}.

The $u$ index denotes the \textit{universality property} :  
the coupling $B$ of the $\ln^2s$ term is the same in all hadron-hadron scatterings and $s_0$ is the same in all reactions.
We note that the coupling $B$ is remarkably stable for the different models : 0.3157~mb (RRPL2$_u$(19)), 0.3152~mb (RRPL2$_u$(21)), 0.3117~mb ((RR)$_c$PL2$_u$(15)), etc. 
This reinforces the validity of the universality property.

We note also that the familiar RRE Donnachie-Landshoff model  is \textit{rejected} at the 98\% C.L. when models which achieve a $\chi^2/dof$ less than 1 for $\sqrt{s}\ge 5$ GeV are considered. 

The predictions of the best RRPL2$_{u}$
model, adjusted for $\sqrt{s} \geq 5$ GeV,  are given in Tables 2-4. 

The uncertainties on total cross sections, including the systematic errors 
due to contradictory data points from FNAL (the CDF and E710/E811 experiments, respectively),
can reach $1.9\%$ at RHIC, $3.1\%$ at the Tevatron, and $4.8\%$ at the LHC,
whereas
those on the $\rho$ parameter are respectively $5.4\%$, $5.2\%$, and $5.4\%$.
The global picture emerging from fits to all data on forward observables
supports 
the CDF data and disfavors the 
E710/E811 data at $\sqrt{s}=1.8$ TeV.

Any significant deviation from the predictions based on model RRPL2$_u$ 
will lead to a re-evaluation of the hierarchy of models and presumably
change the preferred parametrisation to another one. 
A deviation from the ``allowed region" 
would be an indication that strong interactions demand
a generalization of the analytic models discussed so far, {\it e.g.} by adding
Odderon terms, or new Pomeron terms, as suggested by QCD.

\section{A generalization of the Heisenberg model for the total cross-section}

 In his remarkable paper of 1952, Heisenberg investigated production of mesons as a
  problem of shock waves. 
  One of his results was that the total cross section increases like the square of the
  logarithm of the centre-of-mass energy. 
  It is noteworthy that this result coincides with very recent calculations based on
  AdS/CFT dual string-gravity theory \cite{gid}
  or on the Colour Glass Condensate Approach \cite{fiim02} and, of course, saturates
  the Froissart-Martin bound \cite{fro61}.
  In contradistinction to the latter case however the coefficient of the $\ln^2$
  term is an estimate at finite energies and not an asymptotic bound as the one
  obtained by Lukaszuk and Martin \cite{luk67}.

  We have shown that \cite{dos03} by modifications of the original model of Heisenberg
  motivated by the enormous progress of knowledge in the 50 years that passed thence,
  the model yields some general and even some quantitative results which describe the
  data very well.

  The considerations of Heisenberg concerning the total cross section are essentially
  geometrical ones, but the crucial ingredient is that the energy density and not the
  hadronic density is the essential quantity to be taken into account.

Proton-proton collisions are considered in the centre-of-mass system and the energy
$\sqrt{s}$  is supposed to be high enough that Lorentz contraction allows to view
the nucleons as discs.

Interaction takes place only in the overlap region  and the crucial assumption is made that a reaction can only occur if the energy density is high enough in order to create at least a meson pair.

The result of Heisenberg is 
\begin{equation}
\label{2}
\sigma=\frac{\pi}{m^2}\ \ln^2\ \frac{\sqrt{s}}{k_0}.
\end{equation}

We see that implicitely the assumption has been made that if a meson production is 
  energetically possible, it will happen (black disk).
  Of course, Heisenberg was taking the pion mass for the meson mass.
  For the energy of the produced mesons he deduced, in his dynamical considerations,
  assuming interactions of maximal strength, that the energy $k_{0}$ (for two produced mesons) increases only slowly with energy, at any rate not by a power of $s$.
  Therefore the asymptotically leading term in  the cross section is $(\pi/4m^2_\pi)\ln^2s$,
  the coefficient $\pi/4m_\pi^2$ being 1/4 of the Lukaszuk-Martin bound.
  The argument can be extended easily to hadron-hadron scattering in general,
  and therefore we have the result that the coefficient of the $\ln^2s$ term is
  \textit{universal} for all hadron reactions.

There are two obvious necessary modifications of the Heisenberg
  model :
   \begin{itemize}
    \item [1)] If we want to apply it to all kind of hadrons, we have to take care of
    the different hadron sizes, since in the above treatment all sizes are equal
    to $1/m$.
    \item [2)] We have to take into account that {\em direct} pion exchange,
  though being the exchange with the lightest particle, is not relevant at high
  energies.
This is due to the fact that exchanged gluons have spin 1 and pions spin 0. Therefore already in Born approximation gluon exchange dominates at high energies. 
In Regge theory this is manifested by the fact that intercept of the pion is much lower than that of the Pomeron. 
For the mass we rather insert a mass $M$ in the range of the glueball mass instead of the pion mass $m$, since we believe that the high-energy behavior is dominated by gluon 
exchange.
 \end{itemize} 

We then obtain

  \begin{eqnarray}
  \sigma   &=&   \displaystyle\frac{\pi}{4M^2}\ln^2s\nonumber\\
&& +\displaystyle\frac{\pi}{M}\ln s
  \left\{
  (R_1+R_2)+\frac{1}{M}\ln \frac{\alpha}{k_0}
  \right\}\nonumber \\
  &&+\pi(R_1+R_2)^2+\frac{\pi}{M^2}\ln^2\frac{\alpha}{k_0}
  \nonumber \\
  && ~~~~~~~~~~~~~~~+\frac{2\pi}{M}(R_1+R_2)\ln \frac{\alpha}{k_0},
  \label{sigma} 
  \end{eqnarray}
where $0\le\alpha\le 1$.

 We see that the leading $\ln^2s$ term is still universal, but now dominated rather
  by a glueball than by the pion mass.
  Since $R_1$ and $R_2$ are supposed to be of the size of the electromagnetic radii,
  the second term in Eq. (8) will dominate over the
  $\displaystyle\frac{\pi}{M^2}$ term except at high energies, $s\gg k_0^2/\alpha^2$.

  In order to perform a rough numerical estimate, we may insert for the glueball mass
  a value between 1.4 and 1.7 GeV, yielding
  \begin{equation}
  \label{gb }
  \frac{\pi}{4M^2}=0.11-0.16\ \mbox {mb}.
  \end{equation}
 For $R_1$ and $R_2$ we may insert the electromagnetic radii. 
In contrast to Heisenberg, we insert for $k_0$ the minimal energy of two produced particles. 
Since production seems to occur in clusters with mass around 1.3 GeV \cite{gia79}, we can
put $k_0=2.6$ GeV. The value of $\alpha\ (0\le \alpha\le 1)$ might be process dependent. For very small objects ("onia") it might be very small.

In the past, application of the Heisenberg model to the global analyses of the forward hadronic data were performed in \cite{ey88}, but the universality of the leading term was not discussed there. 
This universality was treated by Gershtein and Logunov \cite{gl84}, who made the
assumption, as in the present paper, that the growth of the hadron-hadron total cross-sections is related to resonance production of glueballs.

The COMPETE value of $B$ (see the previous section) corresponds to a mass $M$ of 1 GeV, a bit small for a glueball, but not unreasonable given the crude approximations.
$Z^{HH}$ (where $Z^{HH}=C_1^{HH}+A^{HH}$) are in the right order of magnitude of $R^2$.

 A consequence of the universal $\ln^2 s$ term is that at asymptotic energies all hadron cross sections become equal. 
At finite but high energies the pion and kaon proton cross sections are therefore expected to rise somewhat faster than the nucleon-nucleon cross sections. 
This seems indeed to be indicated by the data.

\begin{figure*}
\includegraphics*[scale=0.8]{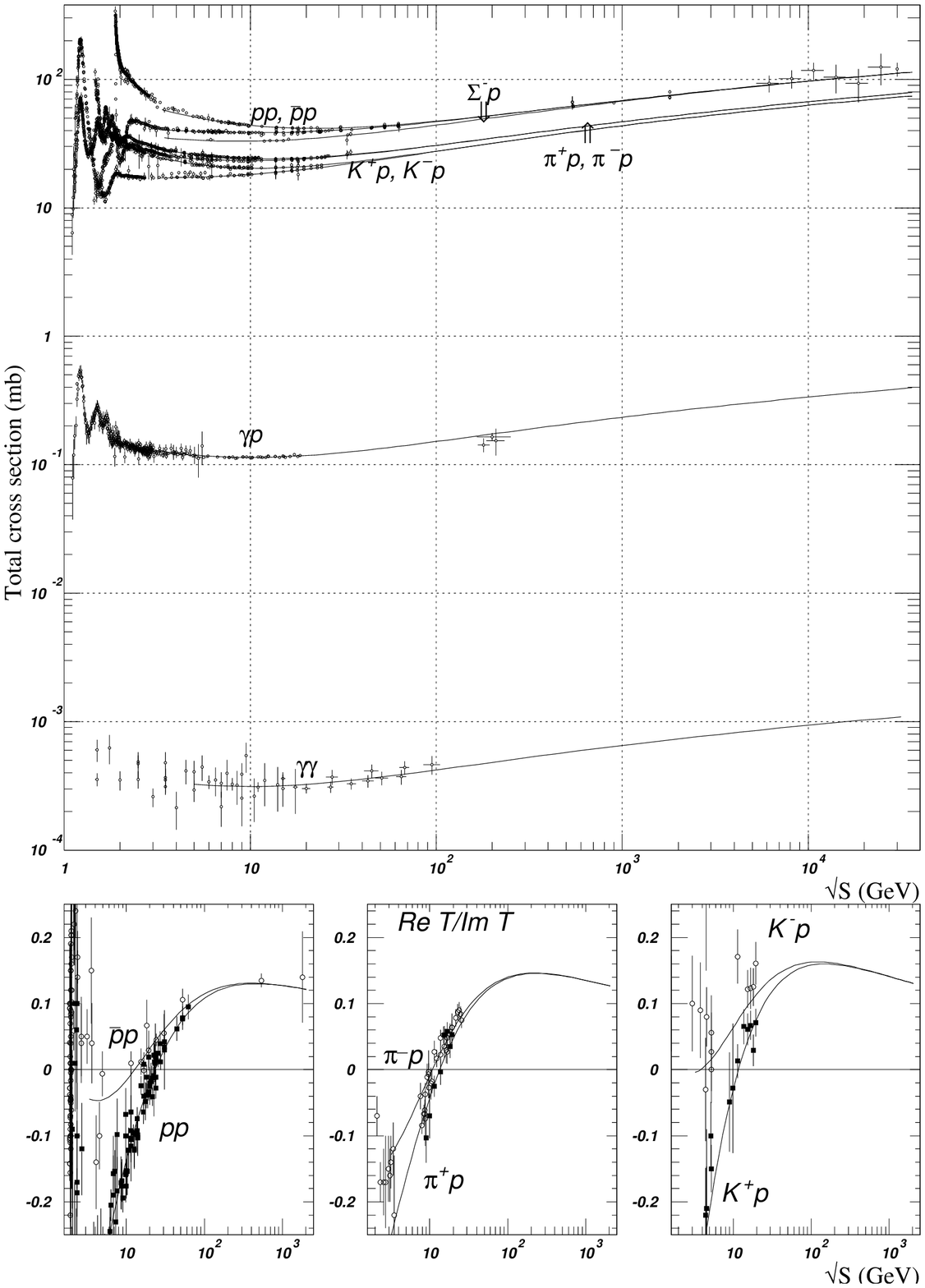}
\caption{Model RRPL2$_u$(19).}
\end{figure*}

\begin{table}[h]
\caption{Ranking of the the 23 models having nonzero area of applicability.
The number between paranthesis (Model Code column) denotes the number of free parameters.}
{\centering \begin{tabular}{lr}
\hline
 Model Code&
 Rank \\
\hline
\hline
 \( {\textrm{RRPL}2_{u}(19)}\)&
 234 \\
\hline
 \( {\textrm{RRP$_{nf}$L}2_{u}(21)} \)&
 221 \\
\hline
 \( {(\textrm{RR}_{c})^{d}\textrm{ PL}2_{u}}(15) \)&
 216 \\
\hline
\( {\textrm{RRL}_{nf}(19)} \)&
 214 \\
\hline
 \( {(\textrm{RR})^{d}\textrm{ P$_{nf}$L}2_{u}}(19) \)&
 206 \\
\hline
 \( {[\textrm{R}^{qc}\textrm{ L}^{qc}]\textrm{R}_{c}}(12) \)&
194 \\
\hline
\( {(\textrm{RR})^{d}\textrm{ PL}2_{u}}(17) \)&
189 \\
\hline
\( {(\textrm{RR}_{c})^{d}\textrm{P}^{qc}\textrm{L}2_{u}}(14) \)&
 185 \\
\hline
 \( {(\textrm{RR})^{d}\textrm{ P$_{nf}$L}2(20)} \)&
182 \\
\hline
\( {(\textrm{RR})^{d}\textrm{P}^{qc}\textrm{L}2_{u}}(16) \)&
 180 \\ 
\hline
\( {\textrm{RR$_c$}L^{qc}(15)} \)&
170 \\
\hline
\( {\textrm{RRL}^{qc}(17)} \)&
164 \\
\hline
 \( {\textrm{RR}_{c}\textrm{ L}2^{qc}(15)} \)&
159 \\
\hline
 \( {\textrm{RRPL}(21)} \)&
155 \\
\hline
 \( {\textrm{RR}\{\textrm{PL2}\}^{qc}}(18) \)&
 155 \\
\hline
 \( {[\textrm{R}^{qc}\textrm{L}^{qc}]\textrm{R}}(14) \)&
154 \\
\hline
 \( {\textrm{RRL}2^{qc}(17)} \)&
153 \\
\hline
 \( {\textrm{RR}\{\textrm{PL2}\}}(20) \)&
 152 \\
\hline
 \( {[\textrm{R}^{qc}\textrm{L}2^{qc}]\textrm{R}_{c}}(12) \)&
 170 \\
\hline
\( {\textrm{RRPE}_{u}}(19) \)&
 146 \\
\hline
 \( {\textrm{RR}_{c}\textrm{PL}}(19) \)&
144 \\
\hline
 \( {\textrm{RRL}2(18)} \)&
 143 \\
\hline
 \( {\textrm{RRL2}(18)} \)&
 142\\
\hline
\end{tabular}\par}
\label{Table1}
\end{table}

\begin{table}[h]
\caption{Predictions for $\sigma_{tot}$ and $\rho$, for $\bar pp$ (at
$\sqrt{s}=1960$ GeV) and for $pp$ (all other energies). The central values 
and statistical errors correspond to the preferred model RRPL2$_u$.}
{\centering \begin{tabular}{ccc}
\( \sqrt{s} \) (GeV)& \( \sigma \) (mb)& \( \rho \)\\
\hline
100& \( 46.37\pm 0.06\)& 
 \( 0.1058\pm 0.0012\)\\
200& \( 51.76\pm 0.12 \)& 
 \( 0.1275\pm 0.0015\)\\
300& \( 55.50\pm 0.17\)&
 \( 0.1352\pm 0.0016\)\\
400& \( 58.41\pm 0.21 \)&
 \( 0.1391\pm 0.0017\)\\
500& \( 60.82\pm 0.25 \)& 
 \( 0.1413\pm 0.0017\)\\
600& \( 62.87\pm 0.28 \)&
 \( 0.1416\pm 0.0018\)\\
1960& \( 78.27\pm 0.55 \)&
 \( 0.1450\pm 0.0018\)\\
10000& \( 105.1\pm 1.1 \)&
 \( 0.1382\pm 0.0016\)\\
12000& \( 108.5\pm 1.2 \)& 
 \( 0.1371\pm 0.0015\)\\
14000& \( 111.5\pm 1.2\)& 
 \( 0.1361\pm 0.0015\)\\
\end{tabular}\par}
\label{table2}
 \end{table}

\begin{table}[h]
\caption{Predictions for $\sigma_{tot}$ for $\gamma p \to hadrons$ for 
cosmic-ray photons. The central values and the statistical errors are as in Table \ref{table2}.}
{\centering \begin{tabular}{cc}
\( p_{lab}^{\gamma} \) (GeV)& \( \sigma \) (mb)\\
\hline
$0.5\cdot10^6$&\( 0.243\pm 0.009\)\\
$1.0\cdot10^6$&\( 0.262\pm 0.010\)\\
$0.5\cdot10^7$&\( 0.311\pm 0.014\)\\
$1.0\cdot10^7$&\( 0.333\pm 0.016\)\\
$1.0\cdot10^8$&\( 0.418\pm 0.022\)\\
$1.0\cdot10^9$&\( 0.516\pm 0.029\)\\
\end{tabular}\par}
\label{table3}
\end{table}

\begin{table}[h]
\caption{Predictions for $\sigma_{tot}$ for $\gamma \gamma \to hadrons$. 
The central values and the statistical errors  are as in Table \ref{table2}.}
{\centering \begin{tabular}{cc}
\( \sqrt{s} \) (GeV)& \( \sigma \) ($\mu$ b)\\
\hline
200&\( 0.546\pm 0.027\)\\
300&\( 0.610\pm 0.035 \)\\
400&\( 0.659\pm 0.042 \)\\
500&\( 0.700\pm 0.047 \)\\
1000&\( 0.840\pm 0.067 \)\\
\end{tabular}\par}
\label{table4}
\end{table}

\end{document}